\documentclass[sigplan,screen]{acmart}
\usepackage{float} 

\acmConference[]{}{}{}

\begin{document}

\title{Open-Source AI-Powered Optimization in Scalene: Advancing Python Performance Profiling with DeepSeek-R1 and LLaMA 3.2}

\author{Saem Hasan}
\authornote{Both authors contributed equally to this research.}
\email{saemhasan@cse.buet.ac.bd}
\affiliation{%
  \institution{Computer Science and Engineering \\Bangladesh University of Engineering and Technology}
  \city{Dhaka}
  \country{Bangladesh}
}

\author{Sanju Basak}
\authornotemark[1]
\email{sanjubasakndc18@gmail.com}
\affiliation{%
\institution{Computer Science and Engineering \\Bangladesh University of Engineering and Technology}
  \city{Dhaka}
  \country{Bangladesh}}

\renewcommand{\shortauthors}{Saem et al.}

\begin{abstract}
Python's flexibility and ease of use come at the cost of performance inefficiencies, requiring developers to rely on profilers to optimize execution. SCALENE, a high-performance CPU, GPU, and memory profiler, provides fine-grained insights into Python applications while running significantly faster than traditional profilers. Originally, SCALENE integrated OpenAI's API to generate AI-powered optimization suggestions, but its reliance on a proprietary API limited accessibility. This study explores the feasibility of using open-source large language models (LLMs), such as DeepSeek-R1 and Llama 3.2, to generate optimization recommendations within SCALENE. Our evaluation reveals that DeepSeek-R1 provides effective code optimizations comparable to proprietary models. We integrate DeepSeek-R1 into SCALENE to automatically analyze performance bottlenecks and suggest improvements, enhancing SCALENE’s utility while maintaining its open-source nature. This study demonstrates that open-source LLMs can be viable alternatives for AI-driven code optimization, paving the way for more accessible and cost-effective performance analysis tools.
\end{abstract}


\maketitle

\section{Introduction}
\begin{figure*}[h]
    \includegraphics[width=\textwidth]{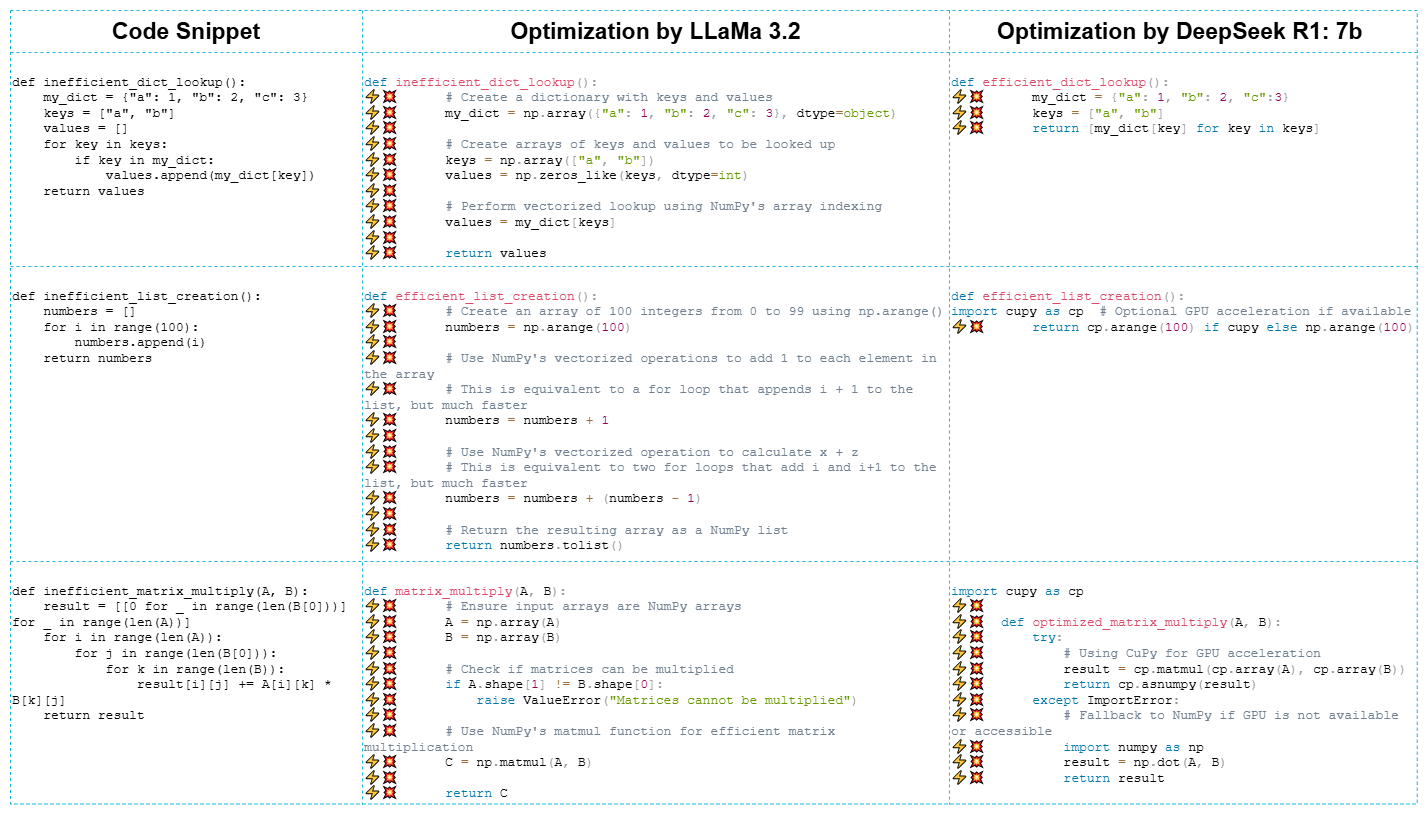} 
    \caption{Some code optimizations suggested by LLaMa3.2 and DeepSeek-R1}
    \label{fig:results}
\end{figure*}
Python’s widespread adoption across industry\cite{python-at-dropbox, python-at-facebook, python-at-instagram, python-at-google, python-at-spotify, python-at-netflix, python-at-youtube, tiobe-index, ieeeplrank2022, redmonk-rankings}, and academia is driven by its ease of use, vast ecosystem, and versatility. However, the very attributes that make Python appealing also introduce significant performance challenges. As a dynamically typed, interpreted language, Python’s standard implementation—CPython—is inherently slower than compiled, native code \cite{enwiki:1095361531}. Because CPython's execution is serialized (it holds a lock whenever it executes a bytecode), it cannot use threads to effectively exploit multiple cores. Despite recent efforts to improve CPython's performance, \footnote{While other other implementations of Python exist, including Jython~\cite{juneau2010definitive} and PyPy~\cite{DBLP:conf/oopsla/RigoP06} CPython is far and away the most popular implementation.} \footnote{For example, the implementation of the interpreter moved from a conventional giant \texttt{switch} statement in version 2.7 to a ``threaded'' interpreter in version 3 that jumps at the end of each byte code implementation to the next, leading to more predictable branches~\cite{10.1145/362248.362270}.} Pure Python code typically runs 1--2 orders
of magnitude slower than native code. As an extreme example, the
Python implementation of matrix-matrix multiplication takes more than
$60,000\times$ as long as the native BLAS version.

Moreover, Python’s high memory overhead, due to the extensive metadata maintained for each object and the nature of its garbage-collected environment, further compounds these inefficiencies. Consequently, even though Python’s rich libraries (such as NumPy\cite{oliphant2006guide}, SciKit-Learn\cite{pedregosa2011scikit}, and TensorFlow\cite{tensorflow2015-whitepaper,abadi2016tensorflow}) enable developers to leverage high-performance native code, much of the code written in pure Python can become a bottleneck in performance-critical applications.

Traditional Python profilers\cite{profile, cprofile, pprofile, memory_profiler} have largely been adapted from tools designed for native code, and they often fall short when addressing Python’s unique execution characteristics. Many of these profilers provide only a limited view—focusing exclusively on either CPU or memory profiling—and typically fail to distinguish between the execution time spent in Python versus native code. This limitation makes it difficult for developers to pinpoint the exact sources of inefficiency and to optimize their programs effectively.

To overcome these challenges, SCALENE\cite{berger2023triangulating} was developed as a groundbreaking profiler specifically tailored for Python. SCALENE combines a suite of innovative techniques to deliver a holistic and fine-grained view of an application’s performance. It simultaneously profiles CPU, memory, and GPU usage, distinguishing between time spent in Python code and native C code. One of its novel contributions is the introduction of a metric known as \emph{copy volume}, which quantifies costly data copying operations that occur when data is moved between Python and native representations or between CPU and GPU memory. These innovations allow SCALENE not only to accurately report performance metrics with minimal overhead but also to reveal subtle inefficiencies that traditional tools miss.

In an effort to further empower developers, SCALENE was extended to provide AI-powered optimization suggestions. Initially, SCALENE leveraged OpenAI’s proprietary API\cite{openai_api} to automatically analyze profiling data and propose targeted code improvements. However, reliance on a paid and closed-source API introduced concerns regarding cost and accessibility. Motivated by these issues, our study explores the integration of open-source language models—specifically DeepSeek-R1\cite{guo2025deepseek} and LLaMA 3.2\cite{meta2024llama}—into the SCALENE framework. We aim to determine whether these open-source models can deliver comparable optimization recommendations without the financial and licensing constraints associated with proprietary solutions.

Through this study, we demonstrate that AI-driven code optimization can be effectively achieved using open-source models integrated into SCALENE, which now offers Ollama options. Specifically, we explored the use of LLaMA 3.2 and DeepSeek-R1 within SCALENE to generate robust optimization recommendations. This approach opens the door to a cost-effective, fully open-source profiling solution that empowers developers to enhance the efficiency of their Python applications without sacrificing the detailed insights provided by modern profiling tools.





\section{Open-Source AI Integration in SCALENE}
As part of our study, we initially sought to integrate open-source large language models (LLMs) directly into SCALENE to enable AI-powered optimization recommendations for Python code. While exploring potential approaches, we discovered that SCALENE now provides built-in support for Ollama\cite{ollama}, an open-source framework that simplifies running LLMs on local machines.  

Ollama is primarily written in Go and serves as a lightweight, efficient runtime for open-source LLMs. Unlike traditional model-serving methods that require complex setup and dependency management, Ollama provides a straightforward API layer, enabling applications like SCALENE to interact seamlessly with models. This API-based integration allows developers to run, query, and manage LLMs locally, eliminating the need for external cloud-based APIs such as OpenAI's proprietary service.  

Recognizing the potential of this approach, we proceeded to download and locally deploy DeepSeek-R1 and LLaMA 3.2—two state-of-the-art open-source models—to explore their effectiveness in generating Python optimization suggestions. By leveraging Ollama’s API, we integrated these models into SCALENE, allowing it to analyze profiling data and propose optimizations based on AI-driven insights.  

This setup enabled us to evaluate the feasibility of using open-source LLMs as a direct replacement for proprietary models in performance profiling. Our findings indicate that DeepSeek-R1, in particular, delivers high-quality optimization suggestions, making it a strong candidate for future AI-driven enhancements in SCALENE. The ability to run such models locally ensures full control over data privacy, removes external API costs, and aligns with SCALENE’s open-source philosophy.  

\section{Experimental Results}

We conducted experiments on 15 different Python code snippets, each covering diverse functionalities such as dictionary lookups, matrix multiplications, and general computational tasks. Among them, three representative examples are presented in Figure \ref{fig:results} to illustrate the comparative effectiveness of DeepSeek R1 and Llama 3.2 in generating optimized code. Our findings indicate that DeepSeek R1 consistently generates more precise and efficient optimizations. It effectively utilizes GPU acceleration via the CuPy library when applicable, significantly improving performance for computationally intensive tasks.

In contrast, Llama 3.2 often produces unnecessary verbose code. While it applies vectorization, it sometimes misinterprets the function’s intent, introducing redundant operations that increase complexity without improving efficiency. This issue is evident in the second example in Figure 1, where it performs unnecessary computations that do not contribute to optimization.

Overall, DeepSeek R1 outperforms Llama 3.2 by providing simpler, computationally efficient, and hardware-aware solutions. Its ability to eliminate redundant computations and leverage GPU acceleration when possible makes it the superior choice for generating optimized Python code. All the code snippets can be found \href{https://docs.google.com/document/d/1I9PEzZMudNeLjdLM6P_H86bxT4WA-9kJpM47QVgBAlc/edit?usp=sharing}{here: http://tiny.cc/scalene-code-sample}.

\section{Conclusion}
The integration of open-source large language models (LLMs) such as DeepSeek-R1 and LLaMA 3.2 into SCALENE marks a significant advancement in Python code optimization. Leveraging Ollama’s framework, SCALENE now delivers AI-driven suggestions locally, reducing reliance on proprietary APIs and lowering operational costs.

Experimental results show that DeepSeek-R1 outperforms LLaMA 3.2 in generating more concise, hardware-aware optimizations. DeepSeek-R1 excels in identifying bottlenecks, eliminating redundant computations, and utilizing GPU acceleration, while LLaMA 3.2 sometimes introduces inefficiencies.

This open-source integration ensures SCALENE remains accessible, transparent, and cost-effective, offering significant performance improvements for Python developers. Future work will focus on expanding supported models and refining optimization suggestions to enhance SCALENE’s versatility.

\bibliographystyle{ACM-Reference-Format}
\bibliography{report}










\end{document}